\documentclass[fleqn,10pt]{wlscirep}
\usepackage[utf8]{inputenc}
\usepackage[T1]{fontenc}

\usepackage{multirow}
\usepackage{hyperref}

\title{Deep learning and whole-brain networks for biomarker discovery: modeling the dynamics of brain fluctuations in resting-state and cognitive tasks}

\author[1,2]{Facundo Roffet}
\author[3,4]{Gustavo Deco}
\author[1,2]{Claudio Delrieux}
\author[3,5,*]{Gustavo Patow}
\affil[1]{Electric and Computer Engineering Department, Universidad Nacional del Sur, Bahía Blanca, Argentina}
\affil[2]{Institute of Computer Science and Engineering, National Scientific and Technological Research Council of Argentina (CONICET), Argentina}
\affil[3]{Center for Brain and Cognition, Computational Neuroscience Group, Department of Information and Communication Technologies, Universitat Pompeu Fabra, Barcelona, Spain}
\affil[4]{Institució Catalana de la Recerca i Estudis Avançats (ICREA), Barcelona, Spain}
\affil[5]{ViRVIG, University of Girona, Girona, Spain}
\affil[*]{gustavo.patow@udg.edu}

\begin{abstract}
{\bf Background}: Brain network models offer insights into brain dynamics, but the utility of model-derived bifurcation parameters as biomarkers remains underexplored. \\ 
{\bf Objective}: This study evaluates bifurcation parameters from a whole-brain network model as biomarkers for distinguishing brain states associated with resting-state and task-based cognitive conditions. \\ 
{\bf Methods}: Synthetic BOLD signals were generated using a supercritical Hopf brain network model to train deep learning models for bifurcation parameter prediction. 
Inference was performed on Human Connectome Project data, including both resting-state and task-based conditions.
Statistical analyses assessed the separability of brain states based on bifurcation parameter distributions.  \\
{\bf Results}: Bifurcation parameter distributions differed significantly across task and resting-state conditions ($p < 0.0001$ for all but two comparisons).
Task-based brain states exhibited higher bifurcation values compared to rest.
At the individual level, a machine learning model was able to classify the predicted bifurcation values into eight cohorts with 62.63\% accuracy (well above the 12.50\% chance level). \\
{\bf Conclusion}: Bifurcation parameters effectively differentiate cognitive and resting states, warranting further investigation as biomarkers for brain state characterization and neurological disorder assessment.  
\end{abstract}
\begin{document}

\flushbottom
\maketitle

\thispagestyle{empty}

\section{Introduction}

Non-invasive neuroimaging techniques are becoming widespread, providing a rich information source that revolutionized the study of brain dynamics, enabling researchers to correlate brain activity with observable behaviors and cognitive processes among other goals~\cite{Fincham2020, Yang2021}.
Functional magnetic resonance imaging (fMRI), in particular, provides data with high spatial resolution detailing oxygen consumption, which is related to regional brain activity.
However, interpreting these data remains a major challenge, as the complexity of brain signals often requires advanced computational approaches and an adequate understanding thereof to uncover meaningful patterns or to elucidate among possible interpretations. 
Deep learning has emerged as a powerful tool for this purpose~\cite{Zhao2018, Thomas2023}, offering flexible models that can learn complex structures directly from raw data without extensive handcrafted features ---they learn directly from the available training samples.
These models excel at extracting intricate patterns across diverse domains and have demonstrated their versatility in a wide range of tasks, including classifying handwritten digits~\cite{Lecun1998}, detecting objects in a scene~\cite{Girshick2015}, segmenting brain tumors~\cite{Havaei2017}, playing board games~\cite{Silver2016}, and bidirectional mapping of structural and functional brain connectivity~\cite{Jamison2024}. 

Biomarkers play a crucial role in advancing 4P medicine (predictive, preventive, personalized, and participatory), enabling, among other things, an early detection of health-compromising conditions (especially those preventable or treatable at an early stage), monitoring disease progression, evaluating the effectiveness of treatments, and tailoring treatments to individual needs.
Given that 4P medicine is inherently participative, it necessitates the collection of extensive and diverse data from a broad range of individuals to ensure comprehensive insights and effective implementation of personalized healthcare strategies.
For this reason, there is an increasing need in neuroscience for adequate and useful biomarkers~\cite{Puig2020}, defined as measurable biological indicators that provide insights to distinguish normal from pathological processes in the brain, or responses to interventions such as therapies or drugs. 
For instance, fMRI-based biomarkers can detect functional connectivity abnormalities, distinguishing Alzheimer’s disease from mild cognitive impairment and healthy controls~\cite{Khatri2022}. 
Similarly, Bruin~et~al.~\cite{Bruin2024} proposed biomarkers to predict outcomes of electroconvulsive therapy, enabling more precise interventions for patients with treatment-resistant depression.

In this context, whole-brain computational models represent a promising approach to understanding the dynamic processes of the brain. 
These models simulate large-scale brain networks, capturing complex neural interactions throughout the brain. 
Whole-brain models have been valuable in various applications, including understanding network-level disruptions in neurological disorders~\cite{Pathak2022}, guiding potential therapeutic interventions~\cite{Deco2014}, and optimizing simulations to achieve high fidelity with limited computational resources~\cite{Herzog2024}. 
An important example is the study by Deco~et~al.~\cite{Deco2017} where the parameters of a simple brain network model were optimized, identifying a \emph{dynamical cortical core} responsible for driving activity across the brain. 
Building on this, the field is moving towards more complex models with high-dimensional, region-specific parameters. Such approaches have been shown to significantly improve the fit to empirical data and enable the prediction of individual traits, though they require dedicated and computationally intensive optimization strategies for each subject \cite{Wischnewski2025}.
While these approaches provided valuable insights, a key unresolved question is whether these model-derived parameters could serve as biomarkers for specific brain states or cognitive functions, thereby limiting their generalizability and broader applicability.

\begin{figure}[t]
\centering
\includegraphics[width=0.95\linewidth]{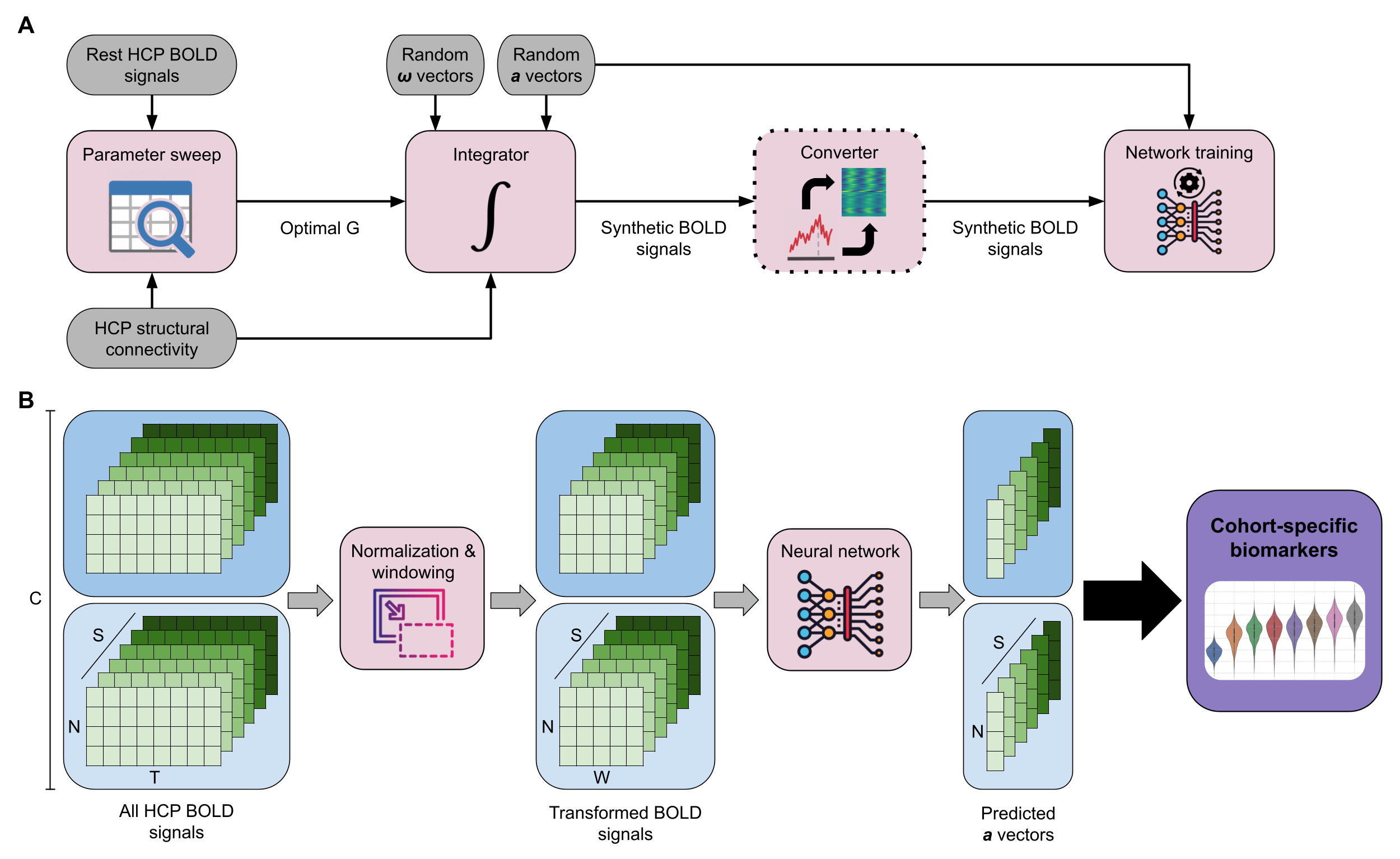}
\caption{\textbf{Overview of the training and inference pipelines for predicting bifurcation parameters from BOLD signals.} 
(A) Synthetic BOLD data is generated using a brain network model based on the supercritical Hopf bifurcation, following a parameter sweep to determine the optimal coupling factor. 
A deep learning model is trained to predict the bifurcation parameters for each brain node using either a time series or an image-based approach. 
The converter block, bordered by a dotted line, is used exclusively in the image-based approach. 
(B) Preprocessed BOLD data from the Human Connectome Project is used to estimate bifurcation parameter values across cohorts. 
After normalization and windowing, BOLD signals are processed through a trained deep learning model to generate predictions. 
These predicted values provide insights into the underlying brain states across different tasks and resting conditions.
The data dimensionality includes $C=8$ cohorts, $S=1,003$ subjects, $T$ scan time steps, and $W=50$ window time steps.
The value of $T$ varies within each scan.\\
Note a key distinction: empirical data is used for selecting the optimal coupling factor and for inference, while synthetic data is used only for model training.}
\label{fig:pipelines}
\end{figure}

To this end, the present study aims to bridge this gap using a whole-brain computational model to explore whether constructive parameters can be biomarkers for different task-based brain states.
In particular, we focus on a {\em neural mass model}, a mathematical framework that simulates the activity of large groups of neurons, which is capable of capturing fluctuations in brain activity and transitions between brain states~\cite{Freyer2011, Freyer2012}.
This model enables the characterization of the brain's dynamical system through bifurcation parameters, which describe the shifts in the brain's stability and oscillatory patterns. 
Our claim is that these bifurcation parameters may be sufficient to characterize distinct task-based brain states associated with various cognitive tasks. 
We trained deep learning models to predict bifurcation parameters using synthetic BOLD signals generated with Hopf's brain network model (see Fig.~\ref{fig:pipelines}). 
By leveraging synthetic data, we overcome the limitations of scarce labeled datasets and ensure that our models are trained on a standardized, well-controlled set of inputs. 
We then applied the trained models to the Human Connectome Project (HCP) dataset to estimate bifurcation parameter distributions across cohorts performing different cognitive tasks. 
By comparing these distributions, we investigate whether bifurcation parameters can serve as meaningful brain state indicators, potentially offering a novel approach to mapping cognitive functions through model-driven biomarkers.

\section{Results}
This study examines brain dynamics across multiple cognitive tasks using data from the HCP, covering resting-state and seven task-based conditions. 
Our methodology involved a three-stage process that separated brain network model calibration, deep learning model training, and inference. The goal was to predict bifurcation parameters from BOLD signals, which serve as a proxy for brain state characteristics by capturing task-related variations in neural activity across brain regions.

First, for model calibration, we used a brain network model based on the supercritical Hopf bifurcation to simulate interactions among 80 cortical and subcortical brain regions, defined by the DK80 parcellation. 
The global coupling factor ($G$) of the brain network model was calibrated by matching the phase functional connectivity dynamics (FCD) of the simulated data with those observed in the empirical HCP resting-state data (preprocessed using the standard HCP pipeline~\cite{Glasser2013, Smith2013}). 
This ensured that our generative brain network model operates in a dynamically realistic regime.

Second, for model training, we overcome the issue of having a limited sample size of empirical data by using the calibrated brain network model to generate a voluminous synthetic BOLD dataset with known ground-truth bifurcation parameters ($a_j$). 
Deep learning models were then trained and validated exclusively on these synthetic data, using an 80/20 split.

Finally, in the inference stage, the fully trained model was applied to the empirical HCP data from all subjects and cohorts. 
This allowed us to perform both a group-level statistical analysis and an individual-level classification with the inferred bifurcation parameters to identify significant differences across tasks and brain networks, thus providing insights into the underlying causal dynamics~\cite{Deco2017} associated with each cognitive task.

\begin{figure}[t]
\centering
\includegraphics[width=\linewidth]{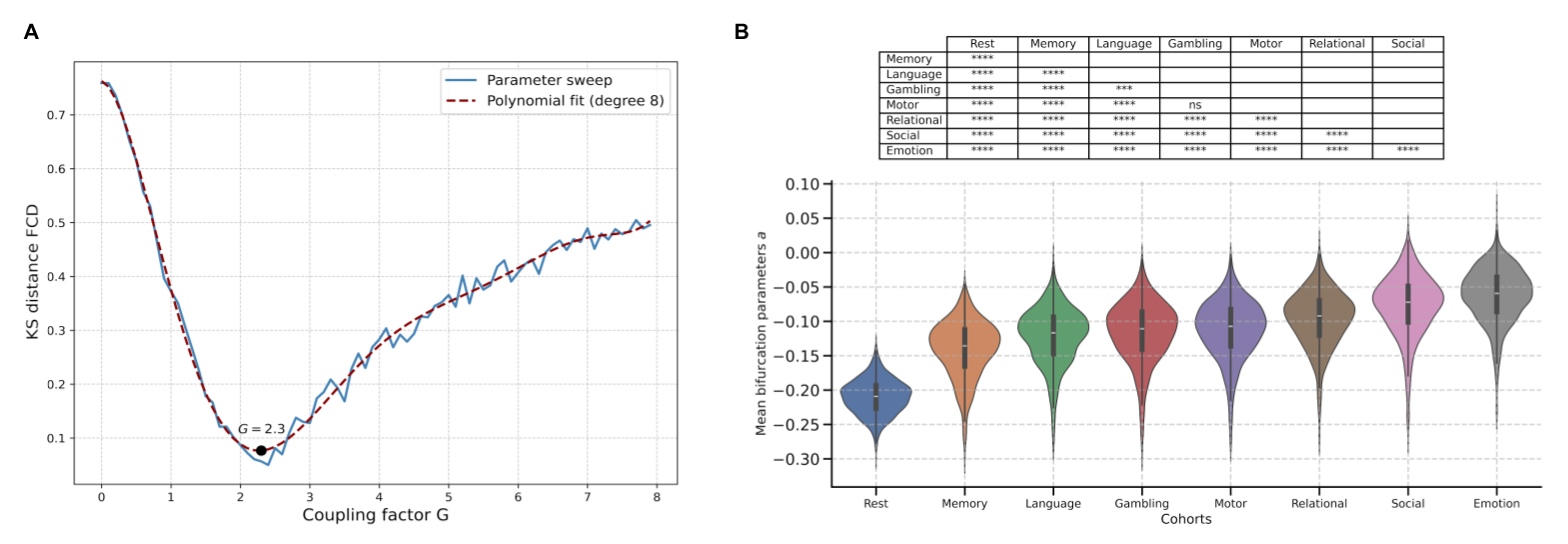}
\caption{\textbf{Determination of optimal coupling factor $G$ and statistical comparison of bifurcation parameters across cohorts.} 
(A) The optimal global coupling factor $G$ was estimated through a parameter sweep, minimizing the Kolmogorov-Smirnov distance between the FCD of empirical and simulated BOLD data. 
The polynomial fit identifies the $G$ value that best aligns the simulated data with the observed resting-state FCD patterns. 
(B) Distribution of mean bifurcation parameter values across different cognitive task and resting-state cohorts.
Statistical significance between cohorts is assessed using the Mann-Whitney-Wilcoxon test, with Benjamini-Hochberg correction applied to control for multiple comparisons. 
Statistical significance is indicated by **** for $p<=0.0001$, *** for $0.0001<p<=0.001$, ** for $0.001<p<=0.01$, * for $0.01<p<=0.05$ and "ns" for $0.05<p<=1$. 
These results reveal distinguishable brain state characteristics associated with each cognitive task.}
\label{fig:results}
\end{figure}

\subsection{Development of the deep learning models}

\subsubsection{Optimal coupling factor $G$}
The optimal fit for the HCP dataset is achieved with an intermediate value of the global coupling factor $G$. 
Specifically, the Kolmogorov-Smirnov (KS) distance between the empirical and simulated FCD is minimized at $G=2.3$ (Fig.~\ref{fig:results}A). 
Since $G$ scales the elements of the structural connectivity matrix $C_{ij}$, normalized to a maximum value of 0.2, this implies that the strongest allowed connection between brain nodes is 0.46.

\subsubsection{Samples are more important than time steps} 
We conducted a comprehensive hyperparameter exploration to understand how the dimensions of the synthetic BOLD signals impact the predictive performance of our deep learning models. 
In this exploration, the number of samples ($S$), nodes ($N$), and time steps ($W$) were jointly adjusted, while ensuring that the total size of the dataset did not exceed 160 million values ($S \times N \times W$) to maintain a limited computational budget. 
Since the selected parcellation fixes the number of nodes at $N=80$, we focused on just varying the values of $S$ and $W$.
For each viable combination of these parameters ({\em i.e.}, those that remained within the imposed threshold), a separate model was trained using the time series approach. The performance metrics of these models are reported in Table~\ref{tab:metrics}.

Our findings confirm that increasing the number of samples consistently improves model performance across all tested configurations. 
This was expected, as more samples generally provide better results when training deep learning models~\cite{Hoffmann2022}. 
However, increasing the number of time steps shows inconsistent improvements. 
Although in some cases a slight boost in performance is noticeable with higher $W$ values, the effect plateaus quickly, and further $W$ increments indeed diminish the performance. 
These results suggest that the most computationally efficient strategy maximizes the number of samples while keeping the time steps relatively low. 
Importantly, we observed a performance degradation when the time step count dropped to $W=25$, indicating that reducing $W$ to this point negatively affects the model's accuracy. 
Based on these insights, we selected $W=50$ and $S=40,000$ as the optimal working point for all subsequent analyses, balancing performance and computational efficiency.

\begin{table}[t]
\centering
\renewcommand{\arraystretch}{1.3} 
\setlength{\tabcolsep}{10pt} 
\begin{tabular}{l|c|cccccc}
\hline
\multirow{2}{*}{} & \multirow{2}{*}{} & \multicolumn{6}{c}{\textbf{Samples ($S$)}} \\ 
\cline{3-8} 
   &                 & \textbf{2,500} & \textbf{5,000} & \textbf{10,000} & \textbf{20,000} & \textbf{40,000} & \textbf{80,000} \\ 
\hline
\multirow{6}{*}{\rotatebox[origin=c]{90}{\textbf{Time steps ($W$)}}} 
   & \textbf{25}  & 14.68 & 13.11 & 12.06 & 11.22 & 10.68 & 10.12 \\ 
   & \textbf{50}  & 13.79 & 12.47 & 11.63 & 10.83 & \textbf{9.82} &        \\ 
   & \textbf{100} & 13.64 & 12.50 & 11.47 & 11.37 &         &        \\ 
   & \textbf{200} & 13.39 & 12.47 & 12.00 &       &         &        \\ 
   & \textbf{400} & 13.65 & 12.47 &       &       &         &        \\ 
   & \textbf{800} & 13.84 &       &       &       &         &        \\ 
\hline
\end{tabular}
\caption{\label{tab:metrics} \textbf{Model performance metrics for different configurations of samples ($S$) and time steps ($W$) in the training dataset.} 
Each model was trained using a unique combination of $S$ and $W$ values, with the total dataset size capped at 160 million values ($S \times N \times W$, with $N=80$ nodes) to ensure computational feasibility.
Model performance is evaluated using a normalized version of the root mean squared error (RMSE), with lower values indicating better predictive accuracy. 
Results indicate that increasing the number of samples leads to more accurate models while increasing time steps has a smaller and inconsistent effect on performance. 
All models reported in this table were trained using the time series approach.}
\end{table}

\subsubsection{The image approach yields better results}
At the optimal working point ($W=50$ and $S=40,000$), we trained an additional model using the image-based approach to predict the bifurcation parameters. 
This approach resulted in a metric value of 6.93, representing a substantial improvement over the 9.82 achieved by the time series approach (see Table~\ref{tab:metrics} for reference). 
Therefore, this model was selected for conducting subsequent analyses.

\subsubsection{Node ordering and parcellation choice impact model performance}
\label{sec:ordering}
To formally test whether the anatomical ordering of brain regions provided a performance advantage, we compared the performance of our original model against a null distribution. 
This null distribution was generated by training the same model ten separate times, each on a unique, randomly permuted ordering of the input nodes.
The models trained on shuffled data achieved a normalized RMSE of 7.12 ± 0.11, with values ranging from 6.97 to 7.35. 
We can compare this to our original metric of 6.93, which was superior to all ten models trained on permuted data. 
To assess statistical significance, we conducted a one-tailed permutation test, which resulted in a p-value of 0.09, the lowest possible value achievable with ten permutations.
To quantify the magnitude of this difference, we also calculated the effect size, which was found to be very large (Cohen's d was 1.76)~\cite{Sullivan2012}.
While the p-value is considered marginally significant due to the limited number of permutations, the large effect size provides strong evidence that the anatomical ordering offers a meaningful, albeit small, performance advantage. 
This indicates that the convolutional network is likely leveraging subtle spatial features in the anatomically ordered data, though its strong performance is not solely dependent on this specific ordering.

In addition, to assess consistency across different brain atlases, we performed a training run using the Schaefer100 parcellation. 
This resulted in a normalized RMSE of 8.53. 
It is important to interpret this result with caution, as the increasing number of nodes likely requires a longer training process to achieve optimal performance. 
However, this result confirms that the choice of parcellation has a tangible impact on the performance of the model and highlights the importance of atlas selection in this framework.

\subsection{Analysis of predicted bifurcation parameters from the HCP dataset}

\subsubsection{Group-level statistical analysis}
Inference on the HCP dataset reveals distinct patterns in the distribution of mean bifurcation parameters across cohorts, supporting their separability based on brain state characteristics. 
As detailed in the Methods section, we conducted pairwise comparisons of these distributions, with the results presented in Fig.~\ref{fig:results}B as violin plots with corresponding Mann-Whitney-Wilcoxon test statistics.  
Remarkably, all pairwise comparisons except one (Motor vs. Gambling) showed statistically significant distribution differences with p-values below 0.001.
The plot further illustrates that most bifurcation parameter values are small negative numbers between -0.25 and 0, indicating a system preference for values in the non-oscillatory behavior of the supercritical Hopf bifurcation~\cite{Deco2017}. 
The resting-state cohort exhibits noticeably lower bifurcation parameter values overall, supporting the hypothesis that these parameters correlate with brain activity. This suggests that greater oscillatory dynamics are required to couple nodes and enhance inter-regional brain communication as the cognitive demands of a task increase.

\subsubsection{Individual-level classification}
\begin{figure}[t]
\centering
\includegraphics[width=0.9\linewidth]{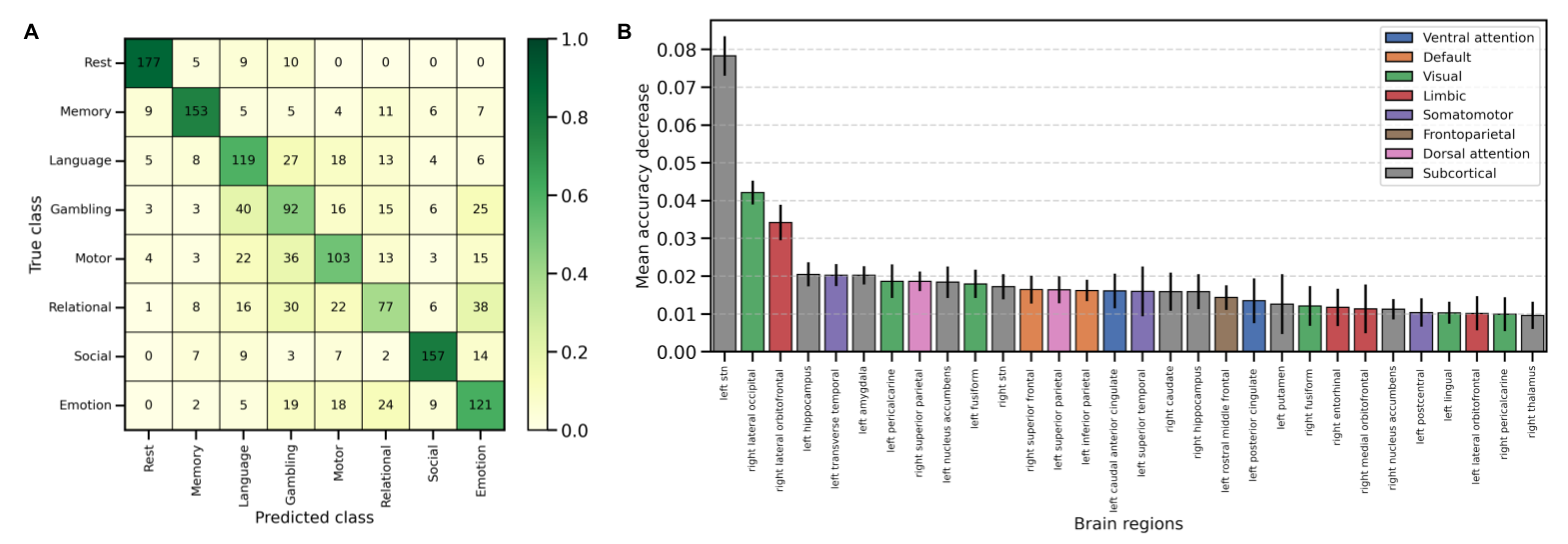}
\caption{\textbf{Individual-level classification of cohorts using bifurcation parameters.}
(A) Confusion matrix of the linear SVM classifier on the held-out test set for the 8-class (seven tasks and rest) classification problem. 
Each cell contains the raw count of predictions. 
The color intensity is normalized by row, representing the recall for each true class (darker green indicates higher recall). 
The diagonal highlights successful classification, while off-diagonal elements show specific confusion patterns between states.
(B) Permutation feature importance of the top 30 most influential brain regions for the classification. 
The y-axis shows the mean decrease in model accuracy after permuting the values for each region 10 times. 
Error bars represent the standard deviation across permutations. 
Brain regions are colored by their corresponding brain networks as defined by the Yeo atlas (also including subcortical).}
\label{fig:classification}
\end{figure}

While group-level analyses revealed significant differences between cohorts, we performed a classification analysis to test whether the inferred bifurcation parameters could serve as effective biomarkers at the individual level. 
A linear Support Vector Machine (SVM) classifier was trained to predict the cognitive state (seven tasks and resting state) for each subject based on their bifurcation parameter vector.
The model demonstrated strong predictive power, achieving a mean accuracy of 62.71\% $\pm$ 2.02\% in a stratified 10-fold cross-validation scheme. 
This is substantially higher than the chance level of 12.50\% for an 8-class problem. 
The performance on a held-out test set was consistent, with an accuracy of 62.63\%. 
The confusion matrix (Fig.~\ref{fig:classification}A) details the classification performance for each cohort showing that the model performed best on the Rest, Memory, and Social cohorts. 
The largest confusions observed align with the overlapping distributions seen in the group-level analysis (Fig.~\ref{fig:results}B).

To identify which brain regions were most critical for this classification, we conducted a permutation feature importance analysis (Fig.~\ref{fig:classification}B).
The results reveal a non-uniform distribution of informative regions across the brain. 
Notably, 10 of the top 30 regions belonged to subcortical nodes, and the only two nodes in the parcellation that belong to the Dorsal Attention network reached the top 15. 
By far the most influential region was the left subthalamic nucleus, followed by the right lateral occipital and the right lateral orbitofrontal cortices. 
In contrast, nodes from the Frontoparietal, Ventral Attention, and Default networks had few appearances on this list, suggesting their bifurcation parameters were less discriminative between the eight cohorts.

\subsubsection{Brain network activations vary across tasks}
\begin{figure}[t]
\centering
\includegraphics[width=0.9\linewidth]{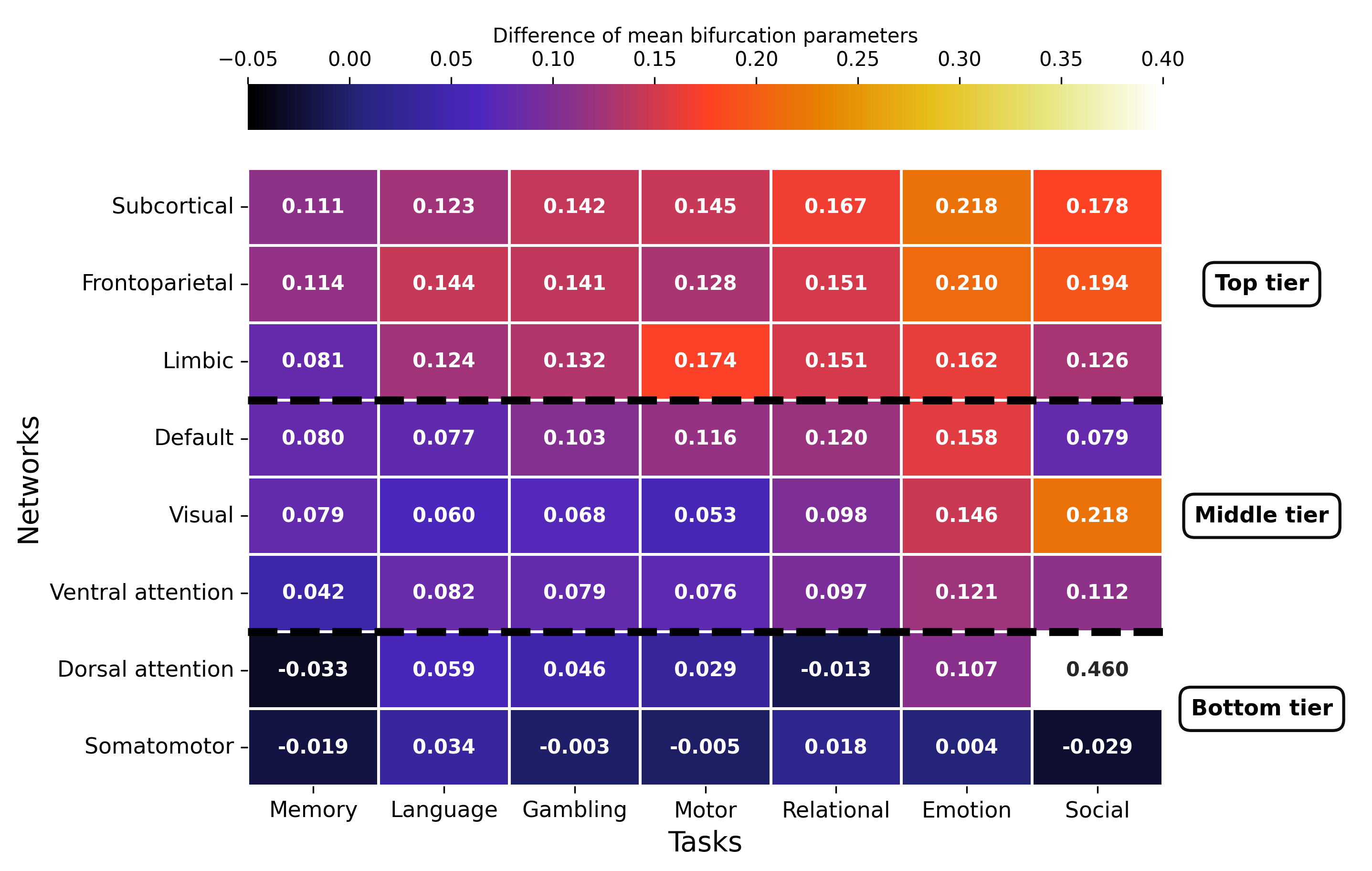}
\caption{\textbf{Task-based differences in mean bifurcation parameters across brain networks.} 
The mean bifurcation parameter values were calculated across cortical nodes grouped by brain network using the Yeo atlas for each task. 
These values were then normalized by subtracting the corresponding mean bifurcation parameter for the resting-state cohort. 
Each cell represents the average difference per network, with colors indicating the magnitude of the difference. Networks are organized by tiers based on overall activation intensity.}
\label{fig:networks}
\end{figure}

To explore how different brain networks are engaged across cognitive functions, we grouped cortical nodes according to their respective brain networks, as defined by the Yeo atlas, and averaged bifurcation parameter values within each cohort.
After subtracting the resting-state values from each task, we obtained the difference of mean bifurcation parameters shown in Fig.~\ref{fig:networks}. 

A primary finding is that nearly all differences are positive, indicating that bifurcation parameters are consistently higher during task engagement compared to rest. This suggests a widespread shift in dynamics across the cortex during active cognition.

Across tasks, we observed a remarkably consistent three-tiered hierarchy of network engagement. The top tier, showing the greatest increase in bifurcation parameters, typically comprised the Subcortical, Frontoparietal and Limbic networks.
A middle tier of moderate activation consisted of the Default, Visual, and Ventral Attention networks. The bottom tier was formed by the Somatomotor network, which showed near-zero changes from rest, and the Dorsal Attention network.
This well-defined hierarchy holds for all conditions except for the Social task, which uniquely disrupts the pattern. In this task, the Dorsal Attention network moves from the bottom tier to become the single most activated network (value of 0.46), showing an exceptionally large increase. The Visual network is also highly activated (0.22), showing a greater response than in any other task. 

Finally, when comparing the overall magnitude of responses, the Memory task induced the smallest changes from rest across most networks. Conversely, the Emotion and Social tasks prompted the largest and most widespread increases in bifurcation parameters, suggesting the highest levels of overall network engagement.

\section{Discussion}

We developed a novel deep-learning-based pipeline that addresses the inverse problem for an ordinary differential equation system, specifically tailored to model BOLD signals.
This approach allows us to infer key dynamical parameters of a computational brain network model with remarkable efficiency and precision, relying solely on synthetic data for model training. 
Our results show that this model-driven inference pipeline can distinguish between different cognitive and resting states in empirical data, providing insights into the underlying brain states and supporting the use of bifurcation parameters as potential biomarkers for neural activity patterns. 
This capability opens a promising landscape for decoding complex task-related brain states and advancing our understanding of human neurocognition in a computationally efficient and scalable manner. 

Our study addresses a significant challenge in neuroscience: large volumes of training data are typically required for deep learning models to generalize effectively~\cite{Perl2020}. 
By generating synthetic BOLD signals derived from a simple whole-brain computational model, we circumvented the limitations associated with scarce or heterogeneous neuroimaging data, offering a controlled, high-volume dataset for training. 
This approach leverages synthetic data to capture essential aspects of BOLD dynamics without reliance on extensive empirical datasets, thereby avoiding potential biases from real-world data collection conditions.
This methodology strengthens the potential of our model to generalize across datasets, highlighting synthetic data as a viable solution in neuroimaging applications where data scarcity often hinders model development.

Previous work in computational brain network modeling has largely focused on examining the dynamics of the human brain in resting states~\cite{Deco2013, Cabral2014, Watanabe2014, Rabuffo2021}. 
In a foundational study, Deco~et~al.~\cite{Deco2017} demonstrated that the brain operates at maximum metastability during rest, characterized by continual transitions between functional network states. 
This metastable behavior is thought to enable optimal integration and segregation of information across regions, facilitating flexible cognitive processing. 
Building on this foundation, our work extends the application of the computational brain network model to task-related brain states, revealing that constructive parameters of the Hopf model ---particularly the bifurcation parameters--- can serve as biomarkers to differentiate distinct mental states induced by various cognitive tasks.
Fig.~\ref{fig:results}B illustrates the increasing demand for inter-regional coordination and dynamic coupling as tasks deviate from the resting-state condition, accentuating the brain's adaptive responses to varying levels of complexity. 
Additionally, the bifurcation parameters we obtained are predominantly small and negative, supporting the claim in this study that the brain operates near the edge of bifurcation.

To the best of our knowledge, this study is among the earliest to adopt an image-based approach for processing BOLD time series using deep learning. 
To date, only Kancharala~et~al. has employed a similar strategy, using Gramian Angular Fields and Markov Transition Fields to encode BOLD signals into image representations~\cite{Kancharala2023}. 
That work achieved an 18\% improvement in classification accuracy when distinguishing brain states induced by viewing images from standard computer vision datasets, compared to conventional time-series-based approaches. 
Analogously, we observed nearly a 3\% decrease in normalized RMSE for bifurcation parameter prediction when using image-based transformations rather than raw time series, underscoring the advantage of this approach. 
By transforming temporal dynamics into spatial representations, image-based methods allow models to leverage powerful computer vision architectures adept at extracting intricate spatial features to capture complex temporal patterns across brain regions. 
Image-based time series methodologies in neuroimaging could pave the way for more robust and accurate tools in functional brain mapping and cognitive state classification.

When analyzing the network-specific task activations shown in Fig.~\ref{fig:networks}, our approach successfully links task-based activation patterns to brain networks commonly associated with these functions, aligning with previous findings in the literature~\cite{Crossley2013}:

\begin{itemize}[nosep]
  \item The Subcortical network, a functionally diverse set of structures including the basal ganglia and thalamus, exhibited consistently high activation across all tasks.
  This is unsurprising given their role as central network hubs that integrate cognitive, motor, and sensory information via extensive cortico-subcortical circuits~\cite{Jansson-Boyd2024}.
  The results reflect their domain-general function in modulating cortical activity and supporting the transition from rest to any computationally demanding cognitive state.
  \item As a core system for executive function~\cite{Niendam2012}, the Frontoparietal network showed high activation across all tasks, peaking during the Emotion and Social tasks. 
  This robust engagement underscores its essential role in integrating information and guiding behavior during demanding cognitive challenges.
  \item The Limbic network, central to emotion and motivation~\cite{Themes2020}, displayed a nuanced activation profile. 
  It was highly engaged during the Motor task, a potentially surprising result suggesting a motivational or affective component to performing the cued movements. 
  As expected, it was also strongly active during the Emotion and Gambling tasks, consistent with its role in emotional processing and reward-based decision-making.
  \item Associated with internal thought~\cite{Greicius2003}, the Default network showed moderate, positive activation across all tasks. 
  This finding challenges a simple view of the network as being purely "task-negative" and supports the modern understanding that it contributes to internal cognitive processing that runs in parallel with external task demands.
  \item The Ventral Attention network facilitates attention shifts and responsiveness to unexpected stimuli~\cite{Bernard2020}. 
  Its moderate activation levels in our study are consistent with tasks that require sustained focus, reflecting the limited role of this network in maintaining attention.
  \item The Dorsal Attention network is responsible for top-down, goal-directed attention, and showed its most dramatic value increase during the Social task. 
  This exceptional activation is likely driven by the need to track the movement and interactions of agents in the video stimuli, a key feature of that specific task design~\cite{Barch2013}. 
  \item The Visual network, which controls sight and pattern recognition, showed activation levels that were clearly modulated by the sensory demands of each task.
  It was lowest during the non-visual Motor and Language tasks, and highest in the Social task which uniquely used video stimuli.
  \item The Somatomotor network, which governs sensory and motor processing~\cite{Biswal1995}, surprisingly presented near-zero differences across all tasks with respect to resting state, even for the Motor task which involves cued finger, toe, and tongue movements.
  Our results may suggest that such structured movements only modulate highly localized sub-regions, an effect that is averaged out across the entire network.
  Alternatively, it could imply that the dynamic state required for these simple, cued actions is not substantially different from the brain's baseline preparedness for movement.

\end{itemize}

Our individual-level classification experiment demonstrates that the inferred bifurcation parameters hold significant predictive power. 
A simple linear model was able to classify a subject's cognitive state with more than 60\% accuracy ---far exceeding chance--- providing strong support for their potential as biomarkers of brain states.
However, the accuracy we achieved with a simple linear SVM does not match the state-of-the-art accuracy for task decoding on the HCP dataset~\cite{Wang2020, Song2022}, limiting immediate clinical application. 
The goal of our analysis was not to compete for classification performance, but rather to validate that model-derived parameters contain meaningful, subject-specific information.

A critical aspect of our methodology is the use of an image-based deep learning model, which proved highly effective. 
But at the same time, our permutation analysis revealed that this effectiveness is, in part, due to the model learning features from the specific spatial ordering of the brain regions in the input matrix. 
This presents a significant limitation: an ideal model should learn the intrinsic dynamics of brain activity irrespective of parcellation choice and arbitrary ordering of nodes. 
Future work should therefore focus on inherently permutation invariant architectures, such as graph neural networks, that can learn the topological properties of the brain network directly, thus creating more robust and generalizable predictive models.
It is also worth noting that the computational brain network model we used is a mesoscopic and phenomenological model, which directly simulates measured BOLD dynamics rather than neural activity. 
This means that regional coupling is applied at the level of hemodynamic signals rather than neural interactions, eliminating the necessity of a convolution with a hemodynamic response function.

The approach developed in this study opens promising venues for the study of bifurcation parameters as biomarkers, with potential applications in assessing brain development, neurodegeneration, and treatment efficacy. 
While fMRI has limited temporal resolution compared to electrophysiological methods like electroencephalography or magnetoencephalography, the framework proposed in this study is modality-agnostic and could be adapted for data with higher temporal sampling.
By establishing the discriminative power of bifurcation parameters, this framework aspires to contribute to the diagnosis and monitoring of neurological conditions.
For instance, since changes in network stability and connectivity are hallmarks of conditions such as Alzheimer’s and Parkinson’s disease~\cite{Damoiseaux2012, Klobusiakova2019}, bifurcation parameters might be used to identify early-stage disruptions in brain dynamics that signal the onset or progression of these diseases.
Future research will focus on individual-level classifications and clinical translation, particularly for monitoring recovery stages in stroke patients.

\section{Methods}

\subsection{Participants and tasks}
This study utilized a sample of 1,003 participants in the March 2017 public release of the Human Connectome Project (HCP). 
Seven distinct cognitive tasks were examined: working memory, motor, gambling, language, social, emotional, and relational ---each designed to engage specific brain regions involved in cognitive and emotional processes.
Detailed task descriptions can be found in Barch~et~al.~\cite{Barch2013}

\subsection{Image acquisition}
All participants were scanned using a Siemens 3-T Connectome-Skyra scanner. 
The scanning protocol was divided into two sessions: the first focused on working memory, gambling, and motor tasks, while the second included language, social cognition, relational processing, and emotion processing tasks. 
In addition to these task-based scans, each participant completed a resting-state scan, during which they viewed a bright cross projected on a dark background for around 15 minutes. 
Comprehensive information about the subjects, scanning protocols, and data preprocessing for both task-based and resting-state sessions is available on the HCP website (\url{http://www.humanconnectome.org/}).

\subsection{Parcellations}
Neuroimaging data preprocessing was performed using a widely adopted atlas, supplemented with subcortical regions to enhance anatomical coverage. 
For a coarser parcellation, the Mindboggle-modified Desikan–Killiany atlas~\cite{Desikan2006} was employed.
This parcellation segmented the cortex into 62 regions, with 31 regions per hemisphere.
Additionally, 18 subcortical regions (nine per hemisphere) were included: hippocampus, amygdala, subthalamic nucleus, globus pallidus internal segment, globus pallidus external segment, putamen, caudate, nucleus accumbens, and thalamus. 
The final parcellation, originally named DK80~\cite{Deco2021}, comprised 80 regions and was aligned with the HCP CIFTI \emph{gray ordinates} standard space to ensure precise regional mapping.
The same parcellation has also been referred to as DBS80 in subsequent works~\cite{Capouskova2022, Alonso2023, Capouskova2023}.
For the cross-parcellation consistency test described in Subsection~\ref{sec:ordering}, the Schaefer100 parcellation~\cite{Schaefer2018} (100 brain regions) was also used.

\subsection{BOLD time series extraction}
\label{sec:ts extraction}
A detailed description of the preprocessing steps applied to the HCP resting-state and task-based datasets is available on the HCP GitHub repository (\url{https://github.com/Washington-University/HCPpipelines)}.
In brief, the HCP preprocessing pipeline was used for resting-state and task-based scans, employing standardized tools such as the FMRIB Software Library, FreeSurfer, and Connectome Workbench~\cite{Glasser2013, Smith2013}. 
The preprocessing steps included: correction for head motion and spatial distortions, intensity normalization, bias field removal, registration to the T1-weighted structural image, and transformation into a 2 mm Montreal Neurological Institute space.
For the resting-state data, the FIX artifact removal procedure was applied. 
Head motion-related noise was regressed, and structured artifacts were removed using independent component analysis (ICA) denoising and the FIX method.
The processed time series for all gray coordinates were projected into the HCP CIFTI \emph{gray ordinates} standard space. 
These files, accessible via surface-based CIFTI formats, were generated for task-based and resting-state conditions. 
To extract the average time series for each brain region defined by the DK80 parcellation, a custom Matlab script was used. 
This script utilized the \texttt{ft\_read\_cifti} function from the Fieldtrip toolbox~\cite{Oostenveld2010}. 
Additionally, the preprocessing pipeline applied a second-order Butterworth filter with a frequency range of 0.008–0.08 Hz to smooth the BOLD signal for both task-based and resting-state datasets. This range is traditionally used to avoid low-frequency drift~\cite{Smith1999} and high-frequency artifacts~\cite{Biswal1995}.

\subsection{Structural connectivity matrix}
The structural connectivity matrix, $C_{ij}$, provides the anatomical backbone for the whole-brain network model. 
It was constructed as a group-average from multi-shell diffusion-weighted imaging data from 32 healthy participants of the Human Connectome Project. 
Tractography was performed using a generalized q-sampling imaging algorithm implemented in DSI Studio. 
For each participant, 200,000 streamlines were seeded within a white-matter mask to reconstruct the brain's fiber tracts. 
The processing pipeline was chosen to minimize the risk of false-positive connections, leveraging a state-of-the-art tracking method that has demonstrated high performance in open challenges~\cite{Maier-Hein2017}. 
The resulting whole-brain connectome was then parcellated using the DK80 atlas to define the connection strength, or fiber density, between each pair of regions.

\subsection{Brain network model}
The brain network model is composed of 80 interconnected brain regions (nodes), identified through a parcellation, as described previously. 
The global dynamics of this network emerge from the interactions among local node dynamics, which are coupled according to the empirically derived anatomical structural connectivity matrix $C_{ij}$ scaled to a maximum value of 0.2.
This value is a conventional choice, since the global coupling factor $G$ serves as the primary parameter to adjust the overall influence of the network on local dynamics.

The local dynamics of each node are governed by the normal form of a supercritical Hopf bifurcation, a model capable of capturing the transition from random, asynchronous behavior to sustained oscillatory activity. 
In Cartesian coordinates, the dynamics of a given node $j$ are described by the following set of equations:
\begin{equation}
\frac{dx_j}{dt} = \left[a_j - x_j^2 - y_j^2\right] x_j - \omega_j y_j + G \sum_i C_{ij} \left(x_i - x_j\right) \beta \eta_j(t),
\end{equation}
\begin{equation}
\frac{dy_j}{dt} = \left[a_j - x_j^2 - y_j^2\right] y_j + \omega_j x_j + G \sum_i C_{ij} \left(y_i - y_j\right) \beta \eta_j(t).
\end{equation}
Here, $\eta_i(t)$ represents additive Gaussian noise with a standard deviation $\beta = 0.02$. 
All synaptic connections in the network are scaled uniformly by a global coupling factor $G$.

This normal form has a supercritical bifurcation at $a_j = 0$, marking a shift in the system's behavior. 
Specifically, for $a_j < 0$, the local dynamics exhibit a stable fixed point at $x_j = 0$ and $y_j = 0$, which, due to additive noise, corresponds to a low-activity asynchronous state. 
Conversely, for $a_j > 0$, the dynamics reach a stable limit cycle oscillation with frequency $f = \omega_j / 2\pi$. 
Thus, the bifurcation parameter $a_j$ serves as a control parameter that defines the activity state of each node. 
By adjusting these parameters, the model can simulate distinct task-related brain states associated with specific cognitive tasks. 
The BOLD signal of each brain region $j$ is then represented by the variable $x_j$, which provides a measurable proxy for neural activity.

The coupling between nodes follows a common difference scheme, which approximates the linear part of a general coupling function. 
This model assumes the weakly coupled oscillator regime, where individual oscillators maintain their intrinsic periodic behavior despite interactions. 
If the linear coupling component vanishes, higher-order nonlinear terms would need to be included, although such cases fall outside the scope of this study.

\subsection{Functional connectivity matrices}
Functional connectivity (FC) quantifies the statistical relationship between brain regions. 
Static FC is computed as the Pearson correlation between the BOLD signals of each pair of brain regions across the entire recording session. 
This process produces an $N \times N$ matrix, where $N$ is the number of nodes, capturing the average spatial organization of brain activity.

Phase functional connectivity dynamics (FCD), on the other hand, assesses how these spatial correlations evolve. 
The procedure to compute FCD is briefly as follows. First, the BOLD signal of each node is transformed using the Hilbert transform to extract its phase.
Then, the cosine of the phase difference between every pair of nodes is used to generate a series of $N \times N$ phase coherence matrices, one for each of the $T$ time points. 
The similarity between these phase coherence matrices is evaluated by calculating the cosine similarity between their upper triangular elements, resulting in a final $T \times T$ matrix that describes the temporal evolution of the FC patterns. 
For further details, readers are encouraged to consult the work of Cabral~et~al.~\cite{Cabral2017}.
When comparing empirical and simulated FCD statistics, the Kolmogorov-Smirnov (KS) distance is used to quantify the largest difference between the cumulative distribution functions of the two samples.

\subsection{Estimation of global coupling factor $G$ and frequencies $\omega$}
Intrinsic frequencies $\omega_j$ were estimated for each brain region from the empirical data by calculating the average peak frequency of the narrowband-filtered BOLD signals for each brain region (see Subsection~\ref{sec:ts extraction}).
The global coupling factor $G$ was determined through an optimization procedure using a parameter sweep, see Fig.~\ref{fig:results}A.
This optimization aimed to minimize the KS distance between the FCDs of the empirical resting-state BOLD signals and those generated by the simulated brain network model. 
Simulations were performed for 20 resting-state subjects, with $G$ values ranging from 0 to 8 in steps of 0.1.
All other model parameters, including the average resting-state structural connectivity matrix $C_{ij}$ and the intrinsic frequency vector \textbf{\emph{$\omega$}} mentioned above, were kept constant. 
The vector \textbf{\emph{a}}, controlling local dynamics, was fixed at $-0.02$ for all nodes, {\em i.e.}, $a_j=-0.02$, ensuring the system operated near the Hopf bifurcation, a critical point where oscillations emerge. 
The resulting curve was smoothed using an eighth-degree polynomial fit after obtaining the KS distance for each $G$ value. 
This polynomial fit allowed for more precise identification of the optimal $G$ value by locating the minimum KS distance.

\subsection{Synthetic data generation}
\label{sec:synthetic data}
Data scarcity poses a particular challenge in neuroscience, as neuroimaging studies are expensive, time-consuming, and often subject to stringent ethical or legal constraints~\cite{Button2013, Sejnowski2014}. 
Apart from this limitation, there are also requirements to control and assess acquisition-related biases, which can vary significantly between scanners, laboratories, and protocols~\cite{Han2006, Brown2011, Roffet2022}.
While the HCP dataset is considered large in neuroscience (1003 scans), it falls short in a deep learning context by at least an order of magnitude.
Synthetic BOLD signals ---closely matching the empirical ones--- were generated to address this issue and properly train a model.

The stochastic Euler-Maruyama method (with $\sigma=0.01$) was used to integrate the system of ordinary differential equations associated with the brain network model, producing $S$ new samples, each consisting of $W$ time steps.
For each sample, the bifurcation parameter $a_j$ values were randomly drawn from a uniform distribution between -1 and 1, and between 0.05 and 0.25 for the intrinsic frequencies $\omega_j$. 
The structural connectivity matrix $C_{ij}$ and the global coupling factor $G$ were constant throughout all simulations.
The initial values of the model variables $x_j$ and $y_j$ were randomly selected from the range $[-1,1]$. 
To ensure that only stable, meaningful dynamics were analyzed, each simulated sample's first 100 time points were discarded to eliminate transient effects.
Thus, if the desired final sample length is $W=50$, the simulation must run for 150 time steps. 
To optimize computational resources during deep learning training, values of $S$ and $W$ were tested, thus determining the most effective configuration (see Table~\ref{tab:metrics}).

\subsection{Deep learning models for predicting bifurcation parameters}
\begin{figure}[t]
\centering
\includegraphics[width=\linewidth]{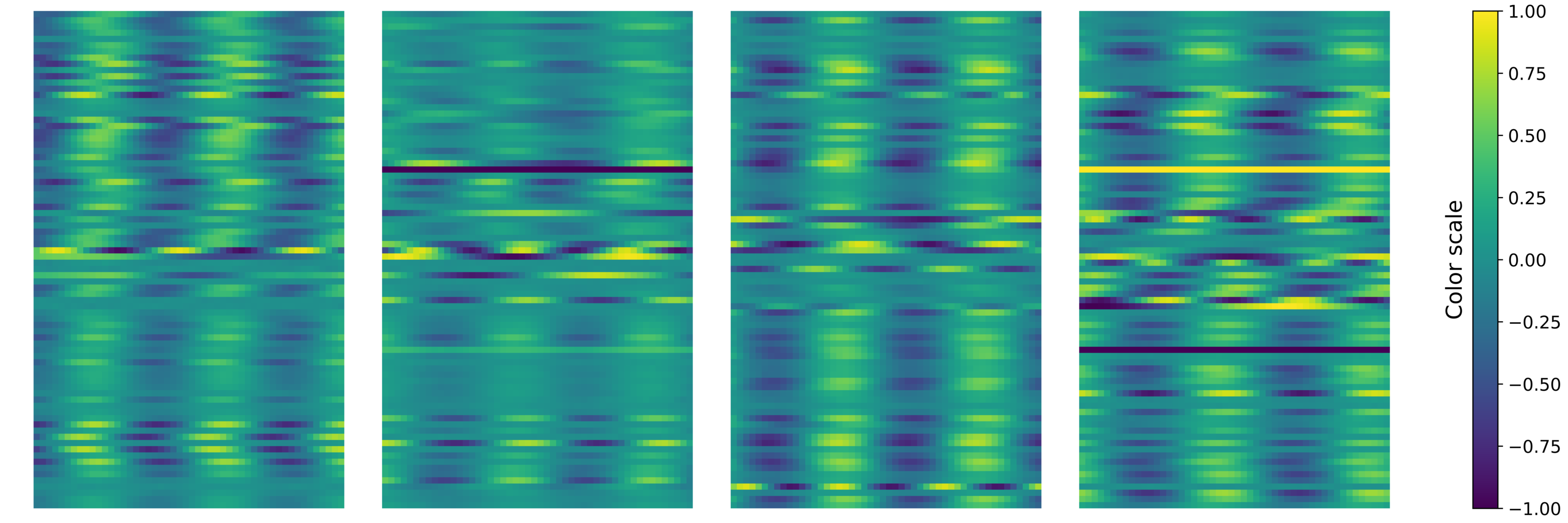}
\caption{\textbf{Examples of BOLD time series converted to images for model input in the image-based approach.} 
Each image represents a single window, with height equal to $N$ (the number of nodes) and width equal to $W$ (the number of time steps chosen). 
The color scale on the right indicates signal values.}
\label{fig:samples}
\end{figure}

Two regression deep learning models were trained exclusively on synthetic data to predict the bifurcation parameters $a_j$ for each node $j$ from a given BOLD signal, as shown in Fig.~\ref{fig:pipelines}A.
The first model follows a standard time-series approach, treating the BOLD signals as sequential data. 
The Temporal Convolutional Network (TCN) architecture~\cite{Bai2018} was used, with the implementation from the \texttt{tsai} Python library (\url{https://github.com/timeseriesAI/tsai}). 
In contrast, the second model adopts a more experimental approach by converting each sample into an image before processing.
This conversion involves mapping the BOLD values to pixels using a convenient color palette, resulting in a picture with a height of $N$ (the number of nodes) and a width of $W$ (the number of time steps). 
Randomly selected samples are displayed in Fig.\ref{fig:samples}. 
This strategy allows to take advantage of current powerful computer vision models, a practice inspired by other deep learning research. 
Time-series-to-image conversion has previously been applied in various fields, such as audio pattern recognition~\cite{Kong2020}, algorithmic financial trading~\cite{Sezer2018}, chaotic system classification~\cite{Uzun2024}, and neuronal activity understanding~\cite{Kancharala2023}. 
In this work, the tiny version of the ConvNeXt architecture~\cite{Liu2022} was used, with the implementation provided by the PyTorch Image Models (\texttt{timm}) library (\url{https://github.com/huggingface/pytorch-image-models}).

Both neural networks were trained using an 80/20 split for training and validation data, a batch size of 16, and a learning rate of 0.0003.
The mean squared error (MSE) loss function was employed with the Adam optimizer over 30 epochs. 
The training process utilized the \texttt{fit\_one\_cycle} method from the \texttt{FastAI} Python library~\cite{Howard2020}. 
The performance of the models was evaluated using a normalized root mean squared error (RMSE) metric, defined as:
\begin{equation}
RMSE_a(\hat{y}, y) = \frac{RMSE(\hat{y}, y) \cdot 100}{a_{max}-a_{min}},
\end{equation}
where $a_{max}=1$ and $a_{min}=-1$, as defined in Subsection~\ref{sec:synthetic data}.

\subsection{Model inference on the HCP dataset}
The HCP BOLD time series were preprocessed before using the trained deep learning models to predict the \textbf{\emph{a}} vectors. 
First, the signals were normalized to a maximum amplitude of 1. 
Next, the time series from the $C$ cohorts and $S$ subjects were divided into non-overlapping windows of $W$ time steps. 
For each cohort, the windows belonging to the same subjects were averaged, resulting in a data dimensionality equal to $C \times S \times W \times N$. 
In the HCP dataset, $S=1003$ and $C=8$ (seven tasks plus rest). 
Since individual scans have varying temporal lengths $T$, this windowing step ensures that each input conforms to the required size of $N \times W$, where $N$ is the number of nodes, and $W$ is the number of time steps used for model training. 
The preprocessed BOLD signals are then fed into the neural network, producing predictions for the bifurcation vector \textbf{\emph{a}}.
The inference pipeline can be visualized in Fig.~\ref{fig:pipelines}B. 
For the image-based model, the time-series-to-image conversion step, described earlier, is also applied before feeding the data into the network.

\subsection{Comparison across cohorts}
The mean $a_j$ values across nodes were calculated to assess differences in bifurcation parameters across cohorts. 
This operation reduced the data dimensionality from $C \times S \times N$ after inference to $C \times S$. 
Statistical differences between distributions were evaluated using the Mann-Whitney-Wilcoxon test, with Benjamini-Hochberg correction applied to control the false discovery rate.

\subsection{Individual-level classification and feature importance}
To assess the discriminative power of the inferred bifurcation parameters at the individual level, we performed a multi-class classification analysis. 
For each subject and cohort, the bifurcation parameter vector was used as a feature vector of length 80.
The HCP data were divided into a training set (80\%) and a held-out test set (20\%) while preserving the proportion of samples for each class. 
A Support Vector Machine (SVM) with a linear kernel was trained on the training data.
The performance of the model was evaluated using a 10-fold stratified cross-validation on the training set, and the final accuracy was reported on the held-out test set.

To determine the contribution of each brain region to the classification, we conducted a permutation feature importance analysis on the trained model using the test set. 
For each of the 80 features (regions), their values were randomly shuffled 10 times, and the resulting decrease in model accuracy was recorded. 
The final importance for each feature was reported as the mean and standard deviation of this decrease in accuracy across the shuffles.

\subsection{Brain network activations}
By interpreting the bifurcation parameters as a proxy for node activation intensity and grouping nodes into brain networks, it is possible to identify which brain regions are generally more active while subjects engage in different tasks. 
In particular, the Yeo atlas~\cite{Yeo2011} divides the cortical surface into seven networks: Default, Visual, Limbic, Somatomotor, Frontoparietal, Ventral Attention, and Dorsal Attention. 
The mean $a$-values across subjects were computed, reducing the data dimensionality from $C \times S \times N$ (cohorts $\times$ subjects $\times$ nodes) to $C \times N$. 
In the case of the DK80 parcellation, $N=80$. 
To account for baseline activity, the resting-state cohort was treated as a reference, and its values were subtracted from the corresponding values of each task, resulting in a matrix of size $(C-1) \times N$. 
Finally, the nodes were grouped into the $B=8$ brain networks (seven Yeo networks plus subcortical nodes), and the values were averaged within each network, yielding a matrix of size $(C-1) \times B$.


\bibliography{bib}

\section*{Data and code availability}
Data used for this research can be downloaded from  \href{http://doi.org/10.5281/zenodo.14508469}{http://doi.org/10.5281/zenodo.14508469}, and code to replicate the results is available on \href{https://github.com/FacuRoffet99/paper-hcp-tasks-hopf-bifurcation}{https://github.com/FacuRoffet99/paper-hcp-tasks-hopf-bifurcation}.

\section*{Funding}
F.R. receives financial support from a doctoral scholarship from the National Scientific and Technical Research Council (CONICET), Argentina. 
G.P. was partially funded by Grant PID2021-122136OB-C22 funded by MCIN/AEI/10.13039/501100011033 and by “ERDF A way of making Europe”, and AGAUR research support grant (ref. 2021 SGR 01035) funded by the Department of Research and Universities of the Generalitat of Catalunya. 
G.D. is supported by Grant PID2022-136216NB-I00 funded by MICIU/AEI/10.13039/501100011033 and by “ERDF A way of making Europe”, ERDF, EU, Project NEurological MEchanismS of Injury, and Sleep-like cellular dynamics (NEMESIS) (ref. 101071900) funded by the EU ERC Synergy Horizon Europe, and AGAUR research support grant (ref. 2021 SGR 00917) funded by the Department of Research and Universities of the Generalitat of Catalunya.

\section*{Author contributions}
F.R., G.P. and G.D. developed the concept. F.R. designed and conducted the experiments. F.R. wrote the first draft of the manuscript. C.D. supervised the research. All authors reviewed and edited the manuscript.

\section*{Competing interests}
The authors declare no competing interests.

\end{document}